\documentclass{article}
\usepackage{amssymb}
\usepackage{amsmath}
\usepackage{float}
\usepackage{graphics}
\usepackage{amsfonts}

\begin{document}

\title{Uncertainty Relation of Functions for Kernel-based Transforms
Obtained in Quantum Mechanics}
\author{Jun-Hua~Chen\thanks{%
Jun-Hua Chen is with $^{1}$Hefei Center for Physical Science and Technology,
Hefei, 230026, China; $^{2}$CAS Key Laboratory of Materials for Energy
Conversion, Hefei, 230026, China; $^{3}$Synergetic Innovation Center of
Quantum Information and Quantum Physics, USTC, Hefei, 230026, China; $^{4}$%
Department of Material Science and Engineering, USTC, Hefei, 230026, China.
email: cjh@ustc.edu.cn } and~Hong-Yi~Fan\thanks{%
Hong-Yi Fan is with $^{1}$Department of Physics, Ningbo University, Ningbo
315211, China; $^{2}$Department of Material Science and Engineering, USTC,
Hefei, 230026, China. email: fhym@ustc.edu.cn}}
\maketitle

\begin{abstract}
In this paper, the generic uncertainty relation (UR) for kernel-based
transformations (KT) of functions is derived. Instead of using the
statistics approach as shown in the literature before, here we employ
quantum mechanical operator approach for directly deriving the UR for KT's.
We are able to do this because we have found the quantum operator
realization of KT. Our new method is concise and applicable to any kinds of
KT's with continuous and discrete parameters and variables. An explicit
result of UR for a family of KT's including FrFT, generalized fractional
transformation (GFrT) and linear canonical transformation (LCT) is provided
as an application of our new method.
\end{abstract}

\section{Introduction}

In optical communication, image manipulation and signal processing, the
fractional Fourier transformation (FrFT) is a very useful tool%
\textsuperscript{\cite{1,2,3,4,5,6,7}}. The concept of the FrFT was
originally described by Condon\textsuperscript{\cite{3}} and later
introduced for signal processing in 1980 by Namias\textsuperscript{\cite{4}}
as a Fourier transform of fractional order. Sumiyoshi et al%
\textsuperscript{\cite{JPA}} also made an interesting generalization on FrFT
in 1994. Working in the context of quantum mechanics (functional analysis),
we have pointed out that any compositable kernel-based transformations $T_{K}%
\left[ f\right] \left( \boldsymbol{B}\right) =\int K\left( \boldsymbol{B},%
\boldsymbol{A}\right) f\left( \boldsymbol{A}\right) d\mu \left( \boldsymbol{A%
}\right) $ can be \textquotedblleft fractionalized" to additive
transformations $T_{\alpha }$ in \cite{CHEN}, where $T_{\alpha }\left[ f%
\right] \left( \boldsymbol{B}\right) =\int K_{\alpha }\left( \boldsymbol{B},%
\boldsymbol{A}\right) f\left( \boldsymbol{A}\right) d\mu \left( \boldsymbol{A%
}\right) $ and $T_{1}=T_{K}$. And we have found the explicit form of all the
compositable and additive kernel-based transformations. We named such
additive transformations as Generalized Fractional Transformation (GFrT).
Since FrFT is a compositable and additive kernel-based transformation, it is
naturally included as one special case of GFrT. The new perspective of
transformations offers many advantages in the calculations in \cite{CHEN},
as we will see later in this work, this new perspective also brings
advantages in dealing with general kernel-based transformations.

On the other hand, the uncertainty principle is always a hot topic in
physics.\textsuperscript{\cite{JMP1,JMP2,JMP3,JMP4}} The uncertainty
principle that describes the constraint on the spreads of functions in the
original domain and transformed domain plays an important role in many
fields like physics, data analysis and signal processing. In \cite{ITE1},
the uncertainty relation (UR) for FrFT on real signals has been calculated
with a large amount of works using traditional method of real analysis. The
UR's for one dimensional linear canonical transformations (LCT) were also
discussed\textsuperscript{\cite{ITE2,ITE3,ITE4}} recently. An interesting
question thus naturally arises: what is the UR for functions undergoing
generalized fractional transformations (GFrT)? And even more generally, what
is the UR for functions undergoing arbitrary kernel-based transformations
(KT)? To our knowledge, only the FrFT and one dimensional LCT had been
concerned regarding the UR of functions in the literatures before.

Instead of employing the usual statistics method (either real analysis
method) to calculate function's variance, in this paper we shall adopt a
completely new approach for deriving the product of the spreads of general
functions in its KT domains with different parameters or even different
types. We are able to accomplish this approach because we have found the
quantum operator realization of KT, thus the whole derivation process can be
carried out in the context of quantum mechanics. The work is arranged as
follows. In Sec. II, we make a brief review of KT in the context of quantum
mechanics. In Sec. III, we convert the calculation of UR for KT to the
related quantum mechanical objects. In Sec. IV, we derive the general UR for
a family of KT, including FrFT, GFrT and multi-dimensional LCT. Then in Sec.
V we apply this formula on four examples to make further illustration of the
method in calculating general UR's for these KT's. The results in \cite%
{ITE1,ITE2,ITE3,ITE4}, which were obtained with huge amount of hard work
there, now appear straightforwardly. A new UR for a complecated KT is also
derived with no special efforts. This is the merit of working in the context
of quantum mechanics.

\section{KT Expressed in the Context of Quantum Mechanics}

Let $\boldsymbol{A=}\left( \boldsymbol{A}_{1},\cdots ,\boldsymbol{A}%
_{m}\right) \in \mathbb{D}_{\boldsymbol{A}}$, $\boldsymbol{B=}\left(
\boldsymbol{B}_{1},\cdots ,\boldsymbol{B}_{n}\right) \in \mathbb{D}_{%
\boldsymbol{B}}$ be $m$ and $n$ dimensional continuous or discrete variables
in Borel sets $\mathbb{D}_{\boldsymbol{A}}$, $\mathbb{D}_{\boldsymbol{B}}$
with measure $\mu \left( \boldsymbol{A}\right) $ and $\mu \left( \boldsymbol{%
B}\right) $ respectively. Since one can always express one complex variable
as two real variables, we can assume that $\boldsymbol{A}$ and $\boldsymbol{B%
}$\ are both real without loss of generality. As is well-known, the
kernel-based transformation $T_{K}$ on function $f$ of $\boldsymbol{A}$ with
kernel function $K\left( \boldsymbol{B},\boldsymbol{A}\right) $ is defined
as the Lebesgue integration
\begin{equation}
T_{K}\left[ f\right] \left( \boldsymbol{B}\right) =\int_{\mathbb{D}_{%
\boldsymbol{A}}}K\left( \boldsymbol{B},\boldsymbol{A}\right) f\left(
\boldsymbol{A}\right) d\mu \left( \boldsymbol{A}\right) .  \label{1}
\end{equation}%
If $\mathbb{D}_{\boldsymbol{A}}=\mathbb{D}_{\boldsymbol{B}}=\mathbb{D}$, and
$\boldsymbol{A}$, $\boldsymbol{B}$ are assigned with the same measure, then
we are able to define the composite transformation $T_{K_{1}}\circ T_{K_{2}}$
of two transformations $T_{K_{1}}\ $and $T_{K_{2}}$ naturally as
\begin{eqnarray}
&&\left( T_{K_{1}}\circ T_{K_{2}}\right) \left[ f\right] \left( \boldsymbol{B%
}\right)  \label{2} \\
&=&\int_{\mathbb{D}}K_{1}\left( \boldsymbol{B},\boldsymbol{A}^{\prime
}\right) K_{2}\left( \boldsymbol{A}^{\prime },\boldsymbol{A}\right) f\left(
\boldsymbol{A}\right) d\mu \left( \boldsymbol{A}\right) d\mu \left(
\boldsymbol{A}^{\prime }\right) .  \notag
\end{eqnarray}%
Such KT's are called \textquotedblleft compositable" KT's. If there exists a
parameterization $K_{\alpha }\left( \boldsymbol{B},\boldsymbol{A}\right) $
of the set of kernels $K\left( \boldsymbol{B},\boldsymbol{A}\right) $'s so
that the composition is additive, i.e.,
\begin{equation}
T_{K_{\alpha }}\circ T_{K_{\beta }}=T_{K_{\alpha +\beta }},  \label{3}
\end{equation}%
then such KT's are called Generalized Fractional Transformation (GFrT). In
other words, GFrT's are additive (and of course compositable, in order to
make sense of additivity) KT's. The details of the construction and the
properties of GFrT are proposed in \cite{CHEN}.

KT's include most of the linear transformations that scientists are
interested in, therefore it is worthwhile to derive a generic UR for
functions undergoing KT's, rather that treating them case by case. KT's are
usually defined by complicated kernels. The complexity of the kernel brings
unnecessary difficulties to the calculations, and makes the meaning of
results obscure.\textsuperscript{\cite{ITE1,ITE2,ITE3,ITE4}} As we will show
later in this work, it helps greatly to simplify the calculations and
clarify the interpretation of the results to work in the context of quantum
mechanics.

In quantum mechanics, a function $f$ is corresponding to a state vector $%
\left\vert f\right\rangle $ in the physics state space (a vector in the
Hilbert space), the value $f\left( \boldsymbol{A}\right) $ of function $f$
at given point $\boldsymbol{A}$ is the inner product $\left\langle
\boldsymbol{A}\right\vert \left. f\right\rangle $. Here \{$\left\vert
\boldsymbol{A}\right\rangle $'s\} and \{$\left\vert \boldsymbol{B}%
\right\rangle $'s\} are two sets of basis of the Hilbert space, satisfying
eigen-equations ($\boldsymbol{\hat{A}=}\left( \boldsymbol{\hat{A}}%
_{1},\cdots ,\boldsymbol{\hat{A}}_{m}\right) $ and $\boldsymbol{\hat{B}=}%
\left( \boldsymbol{\hat{B}}_{1},\cdots ,\boldsymbol{\hat{B}}_{n}\right) $
are some appropriate Hermitian operators)
\begin{equation}
\boldsymbol{\hat{A}}\left\vert \boldsymbol{A}\right\rangle =\boldsymbol{A}%
\left\vert \boldsymbol{A}\right\rangle ,\ \boldsymbol{\hat{B}}\left\vert
\boldsymbol{B}\right\rangle =\boldsymbol{B}\left\vert \boldsymbol{B}%
\right\rangle ,  \label{4}
\end{equation}%
and the completeness relations%
\begin{equation}
\int_{\mathbb{D}_{\boldsymbol{A}}}\left\vert \boldsymbol{A}\right\rangle
\left\langle \boldsymbol{A}\right\vert d\mu \left( \boldsymbol{A}\right)
\boldsymbol{=1},\text{ }\int_{\mathbb{D}_{\boldsymbol{B}}}\left\vert
\boldsymbol{B}\right\rangle \left\langle \boldsymbol{B}\right\vert d\mu
\left( \boldsymbol{B}\right) \boldsymbol{=1}.  \label{5}
\end{equation}%
For example, if the domain of the original function is $\mathbb{R}$, and the
transformed domain is $\mathbb{N}$, then we can choose $\left\vert
\boldsymbol{A}\right\rangle $ to be the 1-dimensional coordinate eigenvector
$\left\vert x\right\rangle $, $\boldsymbol{\hat{A}}=\hat{x}$, and $%
\left\vert \boldsymbol{B}\right\rangle $ to be the photon number eigenstate $%
\left\vert n\right\rangle $, $\boldsymbol{\hat{B}}=\hat{n}$.

Under KT, function $f\left( \boldsymbol{A}\right) =\left\langle \boldsymbol{A%
}\right. \left\vert f\right\rangle $ in $\boldsymbol{A}$ domain is
transformed to new function $T_{K}\left[ f\right] \left( \boldsymbol{B}%
\right) $ in $\boldsymbol{B}$ domain. If $\left\vert \boldsymbol{A}%
\right\rangle $'s and $\left\vert \boldsymbol{B}\right\rangle $'s are chosen
to be orthonormal:
\begin{equation}
\left\langle \boldsymbol{A}\right. \left\vert \boldsymbol{A}^{\prime
}\right\rangle =\delta ^{\left( m\right) }\left( \boldsymbol{A-A}^{\prime
}\right) ,\ \left\langle \boldsymbol{B}\right. \left\vert \boldsymbol{B}%
^{\prime }\right\rangle =\delta ^{\left( n\right) }\left( \boldsymbol{B-B}%
^{\prime }\right) ,  \label{6}
\end{equation}%
then operator $\hat{K}$ that is defined by
\begin{equation}
\hat{K}=\int_{\mathbb{D}_{\boldsymbol{B}}}\int_{\mathbb{D}_{\boldsymbol{A}%
}}K\left( \boldsymbol{B},\boldsymbol{A}\right) \left\vert \boldsymbol{B}%
\right\rangle \left\langle \boldsymbol{A}\right\vert d\mu \left( \boldsymbol{%
A}\right) d\mu \left( \boldsymbol{B}\right)  \label{7}
\end{equation}%
satisfies
\begin{equation}
\begin{array}{c}
\left\langle \boldsymbol{B}\right\vert \hat{K}\left\vert \boldsymbol{A}%
\right\rangle \\
=\int_{\mathbb{D}_{\boldsymbol{B}^{\prime }}}\int_{\mathbb{D}_{\boldsymbol{A}%
^{\prime }}}K\left( \boldsymbol{B}^{\prime },\boldsymbol{A}^{\prime }\right)
\left\langle \boldsymbol{B}\right. \left\vert \boldsymbol{B}^{\prime
}\right\rangle \left\langle \boldsymbol{A}^{\prime }\right. \left\vert
\boldsymbol{A}\right\rangle d\mu \left( \boldsymbol{A}^{\prime }\right) d\mu
\left( \boldsymbol{B}^{\prime }\right) \\
=\int_{\mathbb{D}_{\boldsymbol{B}^{\prime }}}\int_{\mathbb{D}_{\boldsymbol{A}%
^{\prime }}}K\left( \boldsymbol{B}^{\prime },\boldsymbol{A}^{\prime }\right)
\delta ^{\left( m\right) }\left( \boldsymbol{A-A}^{\prime }\right) \delta
^{\left( n\right) }\left( \boldsymbol{B-B}^{\prime }\right) d\mu \left(
\boldsymbol{A}^{\prime }\right) d\mu \left( \boldsymbol{B}^{\prime }\right)
\\
=K\left( \boldsymbol{B},\boldsymbol{A}\right) .%
\end{array}
\label{8}
\end{equation}%
In this case $T_{K}\left[ f\right] \left( \boldsymbol{B}\right) $ can be
rewritten in the context of quantum mechanics as follows
\begin{eqnarray}
T_{K}\left[ f\right] \left( \boldsymbol{B}\right) &=&\int_{\mathbb{D}_{%
\boldsymbol{A}}}K\left( \boldsymbol{B},\boldsymbol{A}\right) f\left(
\boldsymbol{A}\right) d\mu \left( \boldsymbol{A}\right)  \label{9} \\
&=&\int_{\mathbb{D}_{\boldsymbol{A}}}\left\langle \boldsymbol{B}\right\vert
\hat{K}\left\vert \boldsymbol{A}\right\rangle \left\langle \boldsymbol{A}%
\right\vert \left. f\right\rangle d\mu \left( \boldsymbol{A}\right)  \notag
\\
&=&\left\langle \boldsymbol{B}\right\vert \hat{K}\left\vert f\right\rangle ,
\notag
\end{eqnarray}%
where we have used the completeness relation Eq. (\ref{5}) of $\left\vert
\boldsymbol{A}\right\rangle $'s. New expression Eq. (\ref{9}) of KT Eq. (\ref%
{1}) indicates that $T_{K}$ is simply a linear transformation $\hat{K}$ on $%
\left\vert f\right\rangle $ plus the change of basis from $\left\vert
\boldsymbol{A}\right\rangle $ to $\left\vert \boldsymbol{B}\right\rangle $.
Or equivalently, $T_{K}$ is simply the change of basis from $\left\vert
\boldsymbol{A}\right\rangle $ to $\hat{K}^{\dag }\left\vert \boldsymbol{B}%
\right\rangle $ ($\hat{K}^{\dag }$ is the Hermitian conjugation of $\hat{K}$%
), and state $\left\vert f\right\rangle $ is kept unchanged. In the latter
perspective of KT's, performing KT does not change the object $\left\vert
f\right\rangle $ itself. What we do is just choosing different
representations. KT's are naturally passive transformations.

This new perspective of KT's simplifies things greatly. Here are some
examples.

First, in many cases, we demand that the general Parseval's equation holds
for KT, i.e.,

\begin{equation}
\int_{\mathbb{D}_{\boldsymbol{B}}}\left\vert T_{K}\left[ f\right] \left(
\boldsymbol{B}\right) \right\vert ^{2}d\mu \left( \boldsymbol{B}\right)
\equiv \int_{\mathbb{D}_{\boldsymbol{A}}}\left\vert f\left( \boldsymbol{A}%
\right) \right\vert ^{2}d\mu \left( \boldsymbol{A}\right) .  \label{10}
\end{equation}%
Or equivalently%
\begin{equation}
\int_{\mathbb{D}_{\boldsymbol{B}}}T_{K}^{\ast }\left[ f\right] \left(
\boldsymbol{B}\right) T_{K}\left[ g\right] \left( \boldsymbol{B}\right) d\mu
\left( \boldsymbol{B}\right) \equiv \int_{\mathbb{D}_{\boldsymbol{A}%
}}f^{\ast }\left( \boldsymbol{A}\right) g\left( \boldsymbol{A}\right) d\mu
\left( \boldsymbol{A}\right) .  \label{10a}
\end{equation}

In our new perspective, Eq. (\ref{10a}) can be expressed as%
\begin{equation}
\int_{\mathbb{D}_{\boldsymbol{B}}}\left\langle f\right\vert \hat{K}^{\dag
}\left\vert \boldsymbol{B}\right\rangle \left\langle \boldsymbol{B}%
\right\vert \hat{K}\left\vert g\right\rangle d\mu \left( \boldsymbol{B}%
\right) \equiv \int_{\mathbb{D}_{\boldsymbol{A}}}\left\langle f\right.
\left\vert \boldsymbol{A}\right\rangle \left\langle \boldsymbol{A}%
\right\vert \left. g\right\rangle d\mu \left( \boldsymbol{A}\right)
\label{11}
\end{equation}%
Using the completeness relations in Eq. (\ref{5}), Eq. (\ref{11}) becomes
\begin{equation}
\left\langle f\right\vert \hat{K}^{\dag }\hat{K}\left\vert g\right\rangle
\equiv \left\langle f\right. \left\vert g\right\rangle .  \label{12}
\end{equation}%
Since $\left\vert f\right\rangle $ and $\left\vert g\right\rangle $ are
arbitrary states, we have $\hat{K}^{\dag }\hat{K}=$ $\boldsymbol{1}$. In
other words, general Parseval's theorem holds if and only if the KT is
defined by unitary operator $\hat{K}$.

Second, in \cite{CHEN}, we have proved that the kernel of FrFT is in fact%
\begin{eqnarray}
K_{\alpha }\left( p,x\right) &=&\sqrt{\frac{1-i\cot \alpha }{2\pi }}e^{\frac{%
i}{2}\left( \frac{p^{2}+x^{2}}{\tan \alpha }-\frac{2px}{\sin \alpha }\right)
}  \label{13} \\
&=&\left\langle p\right\vert \exp \left[ i\left( \frac{\pi }{2}-\alpha
\right) a^{\dag }a\right] \left\vert x\right\rangle  \notag \\
&=&\left\langle p\right\vert \exp \left[ -i\alpha a^{\dag }a\right]
\left\vert p^{\prime }=x\right\rangle ,  \notag
\end{eqnarray}%
where $\left\vert x\right\rangle $ and $\left\vert p\right\rangle $ are
coordinate and momentum eigenvectors, $a$ and $a^{\dag }$ are the standard
annihilation and creation operator respectively. $\left\vert p^{\prime
}=x\right\rangle $ is a momentum eigenvector with eigenvalue $x$. We see
clearly from Eq. (\ref{13}) that $K_{\pi /2}\left( p,x\right) =$ $%
\left\langle p\right. \left\vert x\right\rangle =\frac{1}{\sqrt{2\pi }}%
e^{-ipx}$ is the traditional FT kernel, and $K_{0}\left( p,x\right)
=\left\langle p\right. \left\vert p^{\prime }=x\right\rangle =\delta \left(
p-x\right) $. The additivity of FrFT is obvious in our new perspective since
\begin{eqnarray}
&&\int K_{\alpha }\left( p,p^{\prime }\right) K_{\beta }\left( p^{\prime
},x\right) dp^{\prime }  \label{14} \\
&=&\int \left\langle p\right\vert \exp \left[ -i\alpha a^{\dag }a\right]
\left\vert p^{\prime }\right\rangle \left\langle p^{\prime }\right\vert \exp %
\left[ -i\beta a^{\dag }a\right] \left\vert p^{\prime \prime
}=x\right\rangle dp^{\prime }  \notag \\
&=&\left\langle p\right\vert \exp \left[ -i\left( \alpha +\beta \right)
a^{\dag }a\right] \left\vert p^{\prime \prime }=x\right\rangle  \notag \\
&=&K_{\alpha +\beta }\left( p,x\right) .  \notag
\end{eqnarray}

Also, as we have shown in \cite{CHEN}, the eigen-problems for GFrT are
simplified greatly in the new perspective.

\section{The UR for KT Derived in the Context of QM}

Under KT, a function $f\left( \boldsymbol{A}\right) $ in $\boldsymbol{A}$
domain is transformed to $T_{K}\left[ f\right] \left( \boldsymbol{B}\right) $
in $\boldsymbol{B}$ domain%
\begin{equation}
f\left( \boldsymbol{A}\right) =\left\langle \boldsymbol{A}\right\vert \left.
f\right\rangle \rightarrow T_{K}\left[ f\right] \left( \boldsymbol{B}\right)
=\left\langle \boldsymbol{B}\right\vert \hat{K}\left\vert f\right\rangle .
\label{15}
\end{equation}%
As usual, the expectation value of $\boldsymbol{\hat{A}}_{i}$ and the
corresponding covariances $Cov\left( \boldsymbol{A}_{i},\boldsymbol{A}%
_{j}\right) $ with respect to signal $f$ are defined as%
\begin{eqnarray}
\boldsymbol{\bar{A}}_{i} &\equiv &\int_{\mathbb{D}_{\boldsymbol{A}}}%
\boldsymbol{A}_{i}\left\vert f\left( \boldsymbol{A}\right) \right\vert
^{2}d\mu \left( \boldsymbol{A}\right)  \label{16} \\
&=&\int_{\mathbb{D}_{\boldsymbol{A}}}\left\langle f\right. \left\vert
\boldsymbol{A}\right\rangle \left\langle \boldsymbol{A}\right\vert
\boldsymbol{\hat{A}}_{i}\left\vert f\right\rangle d\mu \left( \boldsymbol{A}%
\right) =\left\langle f\right\vert \boldsymbol{\hat{A}}_{i}\left\vert
f\right\rangle ,  \notag
\end{eqnarray}%
and%
\begin{eqnarray}
&&Cov\left( \boldsymbol{A}_{i},\boldsymbol{A}_{j}\right)  \label{17} \\
&\equiv &\int_{\mathbb{D}_{\boldsymbol{A}}}\left( \boldsymbol{A}_{i}-%
\boldsymbol{\bar{A}}_{i}\right) \left( \boldsymbol{A}_{j}-\boldsymbol{\bar{A}%
}_{j}\right) \left\vert f\left( \boldsymbol{A}\right) \right\vert ^{2}d\mu
\left( \boldsymbol{A}\right)  \notag \\
&=&\left\langle f\right\vert \left( \boldsymbol{\hat{A}}_{i}-\boldsymbol{%
\bar{A}}_{i}\right) \left( \boldsymbol{\hat{A}}_{j}-\boldsymbol{\bar{A}}%
_{j}\right) \left\vert f\right\rangle ,  \notag
\end{eqnarray}%
where we have used the eigen-equations Eq. (\ref{4}) and the completeness
relation Eq. (\ref{5}). The variance of $\boldsymbol{A}_{i}$ is $\sigma _{%
\boldsymbol{A}_{i}}^{2}=Cov\left( \boldsymbol{A}_{i},\boldsymbol{A}%
_{i}\right) $.

Similarly, we can re-express the average $\boldsymbol{\bar{B}}_{K,i}$ and
covariances $Cov\left( \boldsymbol{B}_{K,i},\boldsymbol{B}_{K,j}\right) $\
in $\boldsymbol{B}$ domain for the transformed signal $T_{K}\left[ f\right]
\left( \boldsymbol{B}\right) $ in the context of quantum mechanics. Because $%
T_{K}\left[ f\right] \left( \boldsymbol{B}\right) =\left\langle \boldsymbol{B%
}\right\vert \hat{K}\left\vert f\right\rangle $, $T_{K}^{\ast }\left[ f%
\right] \left( \boldsymbol{B}\right) =\left\langle f\right\vert \hat{K}%
^{\dag }\left\vert \boldsymbol{B}\right\rangle $ and $\boldsymbol{B}%
_{i}\left\vert \boldsymbol{B}\right\rangle =\boldsymbol{\hat{B}}%
_{i}\left\vert \boldsymbol{B}\right\rangle $, we have%
\begin{eqnarray}
\boldsymbol{\bar{B}}_{K,i} &\equiv &\int_{\mathbb{D}_{\boldsymbol{B}}}%
\boldsymbol{B}_{i}\left\vert T_{K}\left[ f\right] \left( \boldsymbol{B}%
\right) \right\vert ^{2}d\mu \left( \boldsymbol{B}\right)  \label{18} \\
&=&\int_{\mathbb{D}_{\boldsymbol{B}}}\left\langle f\right\vert \hat{K}^{\dag
}\boldsymbol{\hat{B}}_{i}\left\vert \boldsymbol{B}\right\rangle \left\langle
\boldsymbol{B}\right\vert \hat{K}\left\vert f\right\rangle d\mu \left(
\boldsymbol{B}\right)  \notag \\
&=&\left\langle f\right\vert \hat{K}^{\dag }\boldsymbol{\hat{B}}_{i}\hat{K}%
\left\vert f\right\rangle ,  \notag
\end{eqnarray}%
and%
\begin{equation}
\begin{array}{c}
Cov\left( \boldsymbol{B}_{K,i},\boldsymbol{B}_{K,j}\right) \\
\equiv \int_{\mathbb{D}_{\boldsymbol{B}}}\left( \boldsymbol{B}_{i}-%
\boldsymbol{\bar{B}}_{K,i}\right) \left( \boldsymbol{B}_{j}-\boldsymbol{\bar{%
B}}_{K,j}\right) \left\vert T_{K}\left[ f\right] \left( \boldsymbol{B}%
\right) \right\vert ^{2}d\mu \left( \boldsymbol{B}\right) \\
=\left\langle f\right\vert \left( \hat{K}^{\dag }\boldsymbol{\hat{B}}_{i}%
\hat{K}-\boldsymbol{\bar{B}}_{K,i}\right) \left( \hat{K}^{\dag }\boldsymbol{%
\hat{B}}_{j}\hat{K}-\boldsymbol{\bar{B}}_{K,j}\right) \left\vert
f\right\rangle .%
\end{array}
\label{19}
\end{equation}%
The variance of $\boldsymbol{B}_{K,i}$ is $\sigma _{\boldsymbol{B}%
_{K,i}}^{2}=Cov\left( \boldsymbol{B}_{K,i},\boldsymbol{B}_{K,i}\right) $.

From Eqs. (\ref{18}, \ref{19}) we see that the key point to evaluate $%
\boldsymbol{\bar{B}}_{K,i}$ and $Cov\left( \boldsymbol{B}_{K,i},\boldsymbol{B%
}_{K,j}\right) $ is deriving the transformed operator%
\begin{equation}
\boldsymbol{\hat{B}}_{K}\equiv \hat{K}^{\dag }\boldsymbol{\hat{B}}\hat{K}.
\label{20}
\end{equation}%
Once we obtain $\boldsymbol{\hat{B}}_{K}$, then using Eq.(\ref{19}) we can
calculate $Cov\left( \boldsymbol{B}_{K,i},\boldsymbol{B}_{K,j}\right) $ in
any domains. Particularly, we are not constrained to work in $\boldsymbol{B}$
domain.

In the case of GFrT, the operators $\hat{K}$'s can be denoted as $\hat{K}%
_{\alpha }$. We will write $\boldsymbol{\hat{B}}_{K_{\alpha }}$ and $%
\boldsymbol{B}_{K_{\alpha },i}$ as $\boldsymbol{\hat{B}}_{\alpha }$ and $%
\boldsymbol{B}_{\alpha ,i}$ for GFrT. Particularly, when $\alpha =0$,%
\begin{equation}
\boldsymbol{\hat{B}}_{0}=\hat{K}_{0}^{\dag }\boldsymbol{\hat{B}}\hat{K}_{0}=%
\boldsymbol{\hat{A}}.  \label{21}
\end{equation}

For instance, let $\boldsymbol{\hat{A}}$ be the coordinate operator $\hat{X}$%
, $\boldsymbol{\hat{B}}$ the momentum operator $\hat{P}$, the traditional
Heisenberg UR is $\sigma _{X}^{2}\sigma _{P}^{2}\geqslant \frac{1}{4}$. In
FrFT, we have $\hat{K}_{\alpha }=e^{i\left( \frac{\pi }{2}-\alpha \right)
a^{\dag }a}$, then $\hat{K}_{0}=e^{i\frac{\pi }{2}a^{\dag }a}$, $\hat{K}%
_{0}^{\dag }\hat{P}\hat{K}_{0}=\hat{X}$; and $\hat{K}_{\alpha =\frac{\pi }{2}%
}=1$, $\boldsymbol{\hat{B}}_{\frac{\pi }{2}}=\hat{K}_{\frac{\pi }{2}}^{\dag }%
\boldsymbol{\hat{B}}\hat{K}_{\frac{\pi }{2}}=\hat{P}$. $\sigma
_{X}^{2}\sigma _{P}^{2}\geqslant \frac{1}{4}$ can be expressed as $\sigma _{%
\boldsymbol{B}_{0}}^{2}\sigma _{\boldsymbol{B}_{\pi /2}}^{2}\geqslant \frac{1%
}{4}$.

One then naturally asks what is the uncertainty relation for $\sigma _{%
\boldsymbol{B}_{K_{1}}}^{2}$ and $\sigma _{\boldsymbol{B}_{K_{2}}}^{2}$ for
the transformed signals $T_{K_{1}}\left[ f\right] \left( \boldsymbol{B}%
\right) $ and $T_{K_{2}}\left[ f\right] \left( \boldsymbol{B}\right) $
characterized by $K_{1}$ and $K_{2}$ respectively? Or more generally,
suppose we have two KT's characterized by $K_{1}$ and $K_{2}$ and send
signal into different domains $\boldsymbol{B}$ and $\boldsymbol{C}$
respectively, what is the uncertainty relation for $\sigma _{\boldsymbol{B}%
_{K_{1}}}^{2}$ and $\sigma _{\boldsymbol{C}_{K_{2}}}^{2}$?

This problem is quite complicated and tough in statistics\textsuperscript{%
\cite{ITE1,ITE2,ITE3,ITE4}}, and the latter question about $\sigma _{%
\boldsymbol{B}_{K_{1}}}^{2}$ and $\sigma _{\boldsymbol{C}_{K_{2}}}^{2}$ had
not been asked before, to our knowledge. But since we have converted this
problem into the one in the context of quantum mechanics, we can solve it
directly after obtaining operators $\boldsymbol{\hat{B}}_{K_{1}}=\hat{K}%
_{1}^{\dag }\boldsymbol{\hat{B}}\hat{K}_{1}$ and $\boldsymbol{\hat{B}}%
_{K_{2}}=\hat{K}_{2}^{\dag }\boldsymbol{\hat{B}}\hat{K}_{2}$ (or $%
\boldsymbol{\hat{C}}_{K_{2}}=\hat{K}_{2}^{\dag }\boldsymbol{\hat{C}}\hat{K}%
_{2}$). From the knowledge in quantum mechanics, we know that for quantum
state $\left\vert f\right\rangle $ and two Hermitian operators $\hat{U}$ and
$\hat{V}$, there exists the Schr\"{o}dinger-Robertson inequality%
\textsuperscript{\cite{Robertson,Schrodinger}}
\begin{equation}
\left[ \sigma _{\hat{U}}^{2}\right] _{f}\left[ \sigma _{\hat{V}}^{2}\right]
_{f}\geqslant \left\langle \hat{F}\right\rangle _{f}^{2}+\frac{1}{4}%
\left\langle \hat{W}\right\rangle _{f}^{2},  \label{22}
\end{equation}%
where $\left\langle \hat{O}\right\rangle _{f}$ and $\left[ \sigma _{\hat{O}%
}^{2}\right] _{f}$ are the expectation value and the variance of operator $%
\hat{O}$ with respect to the state $\left\vert f\right\rangle $, and
\begin{eqnarray}
\left\langle \hat{F}\right\rangle _{f} &=&\frac{1}{2}\left\langle \hat{U}%
\hat{V}+\hat{V}\hat{U}\right\rangle _{f}-\left\langle \hat{U}\right\rangle
_{f}\left\langle \hat{V}\right\rangle _{f},  \label{23} \\
\hat{W} &=&\frac{1}{i}\left[ \hat{U},\hat{V}\right] .  \notag
\end{eqnarray}

In reference to Eqs. (\ref{18}, \ref{19}, \ref{22}) we immediately have the
UR for KT%
\begin{equation}
\begin{array}{c}
\sigma _{\boldsymbol{B}_{K_{1},i}}^{2}\sigma _{\boldsymbol{B}_{K_{2},j}}^{2}
\\
\geqslant \left\vert \frac{1}{2}\left\langle \left\{ \boldsymbol{\hat{B}}%
_{K_{1},i},\boldsymbol{\hat{B}}_{K_{2},j}\right\} \right\rangle
_{f}-\left\langle \boldsymbol{\hat{B}}_{K_{1},i}\right\rangle
_{f}\left\langle \boldsymbol{\hat{B}}_{K_{2},j}\right\rangle _{f}\right\vert
^{2} \\
+\frac{1}{4}\left\vert \left\langle \left[ \boldsymbol{\hat{B}}_{K_{1},i},%
\boldsymbol{\hat{B}}_{K_{2},j}\right] \right\rangle _{f}\right\vert ^{2}.%
\end{array}
\label{24}
\end{equation}%
Or more generally%
\begin{equation}
\begin{array}{c}
\sigma _{\boldsymbol{B}_{K_{1},i}}^{2}\sigma _{\boldsymbol{C}_{K_{2},j}}^{2}
\\
\geqslant \left\vert \frac{1}{2}\left\langle \left\{ \boldsymbol{\hat{B}}%
_{K_{1},i},\boldsymbol{\hat{C}}_{K_{2},j}\right\} \right\rangle
_{f}-\left\langle \boldsymbol{\hat{B}}_{K_{1},i}\right\rangle
_{f}\left\langle \boldsymbol{\hat{C}}_{K_{2},j}\right\rangle _{f}\right\vert
^{2} \\
+\frac{1}{4}\left\vert \left\langle \left[ \boldsymbol{\hat{B}}_{K_{1},i},%
\boldsymbol{\hat{C}}_{K_{2},j}\right] \right\rangle _{f}\right\vert ^{2}.%
\end{array}
\label{25}
\end{equation}%
Eq. (\ref{25}) provides generic UR's for all KT's, even for two completely
different types of KT's. For example, we can choose the standard Fourier
transform $\mathcal{F}$ as $T_{K_{1}}$, and the decomposition of function as
photonnumber eigenfunctions

\begin{equation}
f\left( x\right) \rightarrow C_{n}=\int \psi _{n}^{\ast }\left( x\right)
f\left( x\right) dx  \label{ex1}
\end{equation}%
as $T_{K_{2}}$. Following Eq. (\ref{25}), we have the UR
\begin{eqnarray}
\sigma _{p}^{2}\sigma _{n}^{2} &\geqslant &\left\vert \frac{1}{2}%
\left\langle \left\{ \hat{p},\hat{n}\right\} \right\rangle _{f}-\left\langle
\hat{p}\right\rangle _{f}\left\langle \hat{n}\right\rangle _{f}\right\vert
^{2}+\frac{1}{4}\left\vert \left\langle \left[ \hat{p},\hat{n}\right]
\right\rangle _{f}\right\vert ^{2}  \label{ex2} \\
&=&\left\vert \frac{1}{2}\overline{\hat{p}\hat{n}+\hat{p}\hat{n}}-\overline{%
\hat{p}}\overline{\hat{n}}\right\vert ^{2}+\frac{1}{4}\overline{\hat{x}}^{2}
\notag
\end{eqnarray}

In the case of GFrT, we have
\begin{equation}
\begin{array}{c}
\sigma _{\boldsymbol{B}_{\alpha ,i}}^{2}\sigma _{\boldsymbol{B}_{\beta
,j}}^{2} \\
\geqslant \left\vert \frac{1}{2}\left\langle \left\{ \boldsymbol{\hat{B}}%
_{\alpha ,i},\boldsymbol{\hat{B}}_{\beta ,j}\right\} \right\rangle
_{f}-\left\langle \boldsymbol{\hat{B}}_{\alpha ,i}\right\rangle
_{f}\left\langle \boldsymbol{\hat{B}}_{\beta ,j}\right\rangle
_{f}\right\vert ^{2} \\
+\frac{1}{4}\left\vert \left\langle \left[ \boldsymbol{\hat{B}}_{\alpha ,i},%
\boldsymbol{\hat{B}}_{\beta ,j}\right] \right\rangle _{f}\right\vert ^{2}.%
\end{array}
\label{26}
\end{equation}%
Thus we have converted the calculation of UR for KT to the related quantum
mechanical operators' commutation relations. In this way the UR's for FrFT
and LCT\textsuperscript{\cite{ITE1,ITE2,ITE3,ITE4}} can be derived briefly
and routinely.

If the operator $\hat{K}$ is known, then the transformed operator $%
\boldsymbol{\hat{B}}_{K}=\hat{K}^{\dag }\boldsymbol{\hat{B}}\hat{K}$ can be
calculated straightforwardly. However in most of the known cases, KT are
defined by $c$-number kernel $K\left( \boldsymbol{B},\boldsymbol{A}\right) $%
. In these cases we are not forced to calculate operator $\hat{K}$. In fact,
we can calculate $\boldsymbol{\hat{B}}_{K}$ directly using $c$-number kernel
$K\left( \boldsymbol{B},\boldsymbol{A}\right) $ as follows,%
\begin{equation}
\begin{array}{c}
\boldsymbol{\hat{B}}_{K}\equiv \hat{K}^{\dag }\boldsymbol{\hat{B}}\hat{K} \\
=\int \left\vert \boldsymbol{A}\right\rangle \left\langle \boldsymbol{A}%
\right\vert \hat{K}^{\dag }\boldsymbol{\hat{B}}\left\vert \boldsymbol{B}%
\right\rangle \left\langle \boldsymbol{B}\right\vert \hat{K}\left\vert
\boldsymbol{A}^{\prime }\right\rangle \left\langle \boldsymbol{A}^{\prime
}\right\vert d\mu \left( \boldsymbol{A}\right) d\mu \left( \boldsymbol{A}%
^{\prime }\right) d\mu \left( \boldsymbol{B}\right) \\
=\int \left[ \boldsymbol{B}K^{\ast }\left( \boldsymbol{B},\boldsymbol{A}%
\right) K\left( \boldsymbol{B},\boldsymbol{A}^{\prime }\right) d\mu \left(
\boldsymbol{B}\right) \right] \left\vert \boldsymbol{A}\right\rangle
\left\langle \boldsymbol{A}^{\prime }\right\vert d\mu \left( \boldsymbol{A}%
\right) d\mu \left( \boldsymbol{A}^{\prime }\right) \\
\equiv \int \boldsymbol{B}_{K}\left( \boldsymbol{A},\boldsymbol{A}^{\prime
}\right) \left\vert \boldsymbol{A}\right\rangle \left\langle \boldsymbol{A}%
^{\prime }\right\vert d\mu \left( \boldsymbol{A}\right) d\mu \left(
\boldsymbol{A}^{\prime }\right) ,%
\end{array}
\label{27}
\end{equation}%
where we have inserted the completeness relations Eq. (\ref{5}) and used the
definition of kernel Eq. (\ref{8}).

Now we have the standard procedure to obtain the UR for given KT. Firstly,
one need to calculate the transformed operators $\boldsymbol{\hat{B}}%
_{K_{1}} $ and $\boldsymbol{\hat{C}}_{K_{2}}$. If the operators $\hat{K}$'s
are not provided explicitly, one can use Eq. (\ref{27}) instead. Secondly,
one need to calculate operators $\left\{ \boldsymbol{\hat{B}}_{K_{1},i},%
\boldsymbol{\hat{C}}_{K_{2},j}\right\} $ and $\left[ \boldsymbol{\hat{B}}%
_{K_{1},i},\boldsymbol{\hat{C}}_{K_{2},j}\right] $. Then we obtain the UR
Eq. (\ref{25}).

The UR's concern the variances $\sigma _{\boldsymbol{B}_{K,i}}^{2}$ of the
transformed function $T_{K}\left[ f\right] \left( \boldsymbol{B}\right) $ in
the new domain $\boldsymbol{B}$. In the original definition of covariance $%
Cov\left( \boldsymbol{B}_{K,i},\boldsymbol{B}_{K,j}\right) =\int_{\mathbb{D}%
_{\boldsymbol{B}}}\left( \boldsymbol{B}_{i}-\boldsymbol{\bar{B}}%
_{K,i}\right) \left( \boldsymbol{B}_{j}-\boldsymbol{\bar{B}}_{K,j}\right)
\left\vert T_{K}\left[ f\right] \left( \boldsymbol{B}\right) \right\vert
^{2}d\mu \left( \boldsymbol{B}\right) $, three objects \textquotedblleft
domain $\boldsymbol{B}$", \textquotedblleft transformation $T_{K}$" and
\textquotedblleft function $f$" are entangled. This entanglement makes
calculations difficult and blurs the meaning of the results. In our new
perspective, \textquotedblleft domain $\boldsymbol{B}$" is represented by
operator $\boldsymbol{\hat{B}}$, \textquotedblleft transformation $T_{K}$"
is represented by operator $\hat{K}$. They are well separated in the new
expression $\left\langle f\right\vert \left( \hat{K}^{\dag }\boldsymbol{\hat{%
B}}_{i}\hat{K}-\boldsymbol{\bar{B}}_{K,i}\right) \left( \hat{K}^{\dag }%
\boldsymbol{\hat{B}}_{j}\hat{K}-\boldsymbol{\bar{B}}_{K,j}\right) \left\vert
f\right\rangle $. All the needed informations are contained in the
transformed operator $\hat{K}^{\dag }\boldsymbol{\hat{B}}\hat{K}$. The
results in \cite{ITE1,ITE2,ITE3,ITE4} are the natural consequences of Eq. (%
\ref{25}) under different situations, which is the result of the Schr\"{o}%
dinger-Robertson inequality Eq. (\ref{22}) and our new perspective of KT's.
If one obtains tighter inequalities compared with Eq. (\ref{22}), then
tighter UR can be obtained immediately for all KT's following our standard
procedure.

In the next section, we will follow the procedure described above to obtain
the UR's for a large family of KT's, including FrFT, GFrT and
multi-dimensional LCT.

\section{UR for a Family of KT including FrFT, GFrT and LCT}

Let $\left\vert \boldsymbol{A}\right\rangle =\left\vert \vec{x}\right\rangle
$ be the $n$-dimensional coordinate representation, and $\left\vert
\boldsymbol{B}\right\rangle =\left\vert \vec{p}\right\rangle $ be the
momentum representation. $\boldsymbol{\hat{A}}=\hat{X}=\left( \hat{X}%
_{1},\ldots ,\hat{X}_{n}\right) ^{T}$ and $\boldsymbol{\hat{B}}=\hat{P}%
=\left( \hat{P}_{1},\ldots ,\hat{P}_{n}\right) ^{T}$ are $n$-dimensional
coordinate and momentum operators. Here superscript $T$ means transpose
operation on matrices. In the following context, we consider KT's whose
kernel take the form

\begin{equation}
K\left( \vec{p},\vec{x}\right) =\frac{\exp \left[ i\left( A_{K}\left( \vec{p}%
\right) +B_{K}\left( \vec{x}\right) -\vec{p}\cdot C_{K}\vec{x}\right) \right]
}{\left( 2\pi \right) ^{n/2}\sqrt{D_{K}}},  \label{28}
\end{equation}%
where $A_{K}\left( \vec{p}\right) $, $B_{K}\left( \vec{x}\right) $\ are
real-value functions of $\vec{p}$, $\vec{x}$ respectively, and $C_{K}$ is an
$n\times n$ nonsingular real matrix. Transformations that satisfy Parseval's
theorem are much more important in physics and signal processing. Therefore
in the following context we consider only such transformations. Parseval's
theorem demands that $\left\vert D_{K}\det C_{K}\right\vert \equiv 1$.

Using Eq. (\ref{27}), the transformed momentum operator is%
\begin{equation}
\begin{array}{c}
\hat{P}_{K}=\int \vec{p}K^{\ast }\left( \vec{p},\vec{x}\right) K\left( \vec{p%
},\vec{x}^{\prime }\right) d^{n}\vec{p}\left\vert \vec{x}\right\rangle
\left\langle \vec{x}^{\prime }\right\vert d^{n}\vec{x}d^{n}\vec{x}^{\prime }
\\
=\frac{1}{\left( 2\pi \right) ^{n}\left\vert D_{K}\right\vert }\int \vec{p}%
e^{i\vec{p}\cdot C_{K}\left( \vec{x}-\vec{x}^{\prime }\right) }d^{n}\vec{p}%
\left\vert \vec{x}\right\rangle \left\langle \vec{x}^{\prime }\right\vert \\
e^{i\left( B_{K}\left( \vec{x}^{\prime }\right) -B_{K}\left( \vec{x}\right)
\right) }d^{n}\vec{x}d^{n}\vec{x}^{\prime } \\
\overset{\vec{p}^{\prime }=C_{K}^{T}\vec{p}}{=}\frac{1}{\left( 2\pi \right)
^{n}}\int \left( C_{K}^{T}\right) ^{-1}\vec{p}^{\prime }e^{i\vec{p}^{\prime
}\cdot \left( \vec{x}-\vec{x}^{\prime }\right) }d^{n}\vec{p}^{\prime
}\left\vert \vec{x}\right\rangle \left\langle \vec{x}^{\prime }\right\vert
\\
e^{i\left( B_{K}\left( \vec{x}^{\prime }\right) -B_{K}\left( \vec{x}\right)
\right) }d^{n}\vec{x}d^{n}\vec{x}^{\prime } \\
=\frac{1}{\left( 2\pi \right) ^{n}}\int \frac{1}{i}\left( C_{K}^{T}\right)
^{-1}\nabla _{\vec{x}}\left[ \int e^{i\vec{p}^{\prime }\cdot \left( \vec{x}-%
\vec{x}^{\prime }\right) }d^{n}\vec{p}^{\prime }\right] \\
\left\vert \vec{x}\right\rangle \left\langle \vec{x}^{\prime }\right\vert
e^{i\left( B_{K}\left( \vec{x}^{\prime }\right) -B_{K}\left( \vec{x}\right)
\right) }d^{n}\vec{x}d^{n}\vec{x}^{\prime } \\
=\frac{1}{i}\int \left( C_{K}^{T}\right) ^{-1}\nabla _{\vec{x}}\left[ \delta
^{\left( n\right) }\left( \vec{x}-\vec{x}^{\prime }\right) \right] \\
\left\vert \vec{x}\right\rangle \left\langle \vec{x}^{\prime }\right\vert
e^{i\left( B_{K}\left( \vec{x}^{\prime }\right) -B_{K}\left( \vec{x}\right)
\right) }d^{n}\vec{x}d^{n}\vec{x}^{\prime }%
\end{array}
\label{29}
\end{equation}%
Integrating by parts in Eq. (\ref{29}), we have finally%
\begin{equation}
\begin{array}{c}
\hat{P}_{K}=i\int \delta ^{\left( n\right) }\left( \vec{x}-\vec{x}^{\prime
}\right) \left( C_{K}^{T}\right) ^{-1} \\
\nabla _{\vec{x}}\left[ \left\vert \vec{x}\right\rangle \left\langle \vec{x}%
^{\prime }\right\vert e^{i\left( B_{K}\left( \vec{x}^{\prime }\right)
-B_{K}\left( \vec{x}\right) \right) }\right] d^{n}\vec{x}d^{n}\vec{x}%
^{\prime } \\
=\int \left( C_{K}^{T}\right) ^{-1}\left[ \left( i\nabla _{\vec{x}%
}\left\vert \vec{x}\right\rangle \right) +\nabla _{\vec{x}}B_{K}\left( \vec{x%
}\right) \left\vert \vec{x}\right\rangle \right] \\
\left\langle \vec{x}^{\prime }\right\vert e^{i\left( B_{K}\left( \vec{x}%
^{\prime }\right) -B_{K}\left( \vec{x}\right) \right) }\delta ^{\left(
n\right) }\left( \vec{x}-\vec{x}^{\prime }\right) d^{n}\vec{x}d^{n}\vec{x}%
^{\prime } \\
=\int \delta ^{\left( n\right) }\left( \vec{x}-\vec{x}^{\prime }\right)
\left( C_{K}^{T}\right) ^{-1}\left( \hat{P}+\nabla _{\hat{X}}B_{K}\left(
\hat{X}\right) \right) \\
\left\vert \vec{x}\right\rangle \left\langle \vec{x}^{\prime }\right\vert
e^{i\left( B_{K}\left( \vec{x}^{\prime }\right) -B_{K}\left( \vec{x}\right)
\right) }d^{n}\vec{x}d^{n}\vec{x}^{\prime } \\
=\left( C_{K}^{T}\right) ^{-1}\left( \hat{P}+\nabla _{\hat{X}}B_{K}\left(
\hat{X}\right) \right) ,%
\end{array}
\label{30}
\end{equation}%
where we have applied the identity $i\nabla _{\vec{x}}\left\vert \vec{x}%
\right\rangle =\hat{P}\left\vert \vec{x}\right\rangle $.

Particularly, when $B_{K}\left( \vec{x}\right) =\frac{1}{2}\vec{x}\cdot
\tilde{B}_{K}\vec{x}$ is a quadratic of $\vec{x}$, where $\tilde{B}_{K}$ is
a real symmetric matrix,
\begin{equation}
K\left( \vec{p},\vec{x}\right) =\frac{\exp \left[ i\left( A_{K}\left( \vec{p}%
\right) +\frac{1}{2}\vec{x}\cdot \tilde{B}_{K}\vec{x}-\vec{p}\cdot C_{K}\vec{%
x}\right) \right] }{\left( 2\pi \right) ^{n/2}\sqrt{D_{K}}},  \label{31}
\end{equation}%
we have $\nabla _{\vec{x}}B_{K}\left( \vec{x}\right) =\tilde{B}_{K}\vec{x}$
and
\begin{equation}
\hat{P}_{K}=\left( C_{K}^{T}\right) ^{-1}\left( \hat{P}+\tilde{B}_{K}\hat{X}%
\right) .  \label{32}
\end{equation}

We immediately have the commutation relations%
\begin{eqnarray}
&&\frac{1}{i}\left[ \hat{P}_{K_{1},i},\hat{P}_{K_{2},j}\right]  \label{33} \\
&=&\left[ \left( C_{K_{1}}^{T}\right) ^{-1}\left( \tilde{B}_{K_{1}}-\tilde{B}%
_{K_{2}}\right) C_{K_{2}}^{-1}\right] _{ij}  \notag \\
&\equiv &W_{K_{1}K_{2},ij}  \notag
\end{eqnarray}%
and%
\begin{equation}
\begin{array}{c}
\frac{1}{2}\left\langle \left\{ \hat{P}_{K_{1},i},\hat{P}_{K_{2},j}\right\}
\right\rangle _{f}-\left\langle \hat{P}_{K_{1},i}\right\rangle
_{f}\left\langle \hat{P}_{K_{2},j}\right\rangle _{f} \\
=\left[ \left( C_{K_{1}}^{T}\right) ^{-1}\left(
\begin{array}{c}
\tilde{B}_{K_{1}}\Delta ^{XP}+\Delta ^{PX}\tilde{B}_{K_{2}} \\
+\tilde{B}_{K_{1}}\Delta ^{XX}\tilde{B}_{K_{2}}+\Delta ^{PP}%
\end{array}%
\right) C_{K_{2}}^{-1}\right] _{ij} \\
\equiv F_{K_{1}K_{2},ij}\ ,%
\end{array}
\label{34}
\end{equation}%
where real matrices $\Delta ^{PP}$, $\Delta ^{PX}$, $\Delta ^{XP}$ and $%
\Delta ^{XX}$ are defined by their elements%
\begin{equation}
\begin{array}{c}
\Delta _{st}^{PP}=\frac{1}{2}\left\langle \left\{ \hat{P}_{s},\hat{P}%
_{t}\right\} \right\rangle _{f}-\left\langle \hat{P}_{s}\right\rangle
_{f}\left\langle \hat{P}_{t}\right\rangle _{f}\ , \\
\Delta _{st}^{PX}=\frac{1}{2}\left\langle \left\{ \hat{P}_{s},\hat{X}%
_{t}\right\} \right\rangle _{f}-\left\langle \hat{P}_{s}\right\rangle
_{f}\left\langle \hat{X}_{t}\right\rangle _{f}\ , \\
\Delta _{st}^{XP}=\frac{1}{2}\left\langle \left\{ \hat{X}_{s},\hat{P}%
_{t}\right\} \right\rangle _{f}-\left\langle \hat{X}_{s}\right\rangle
_{f}\left\langle \hat{P}_{t}\right\rangle _{f}\ , \\
\Delta _{st}^{XX}=\frac{1}{2}\left\langle \left\{ \hat{X}_{s},\hat{X}%
_{t}\right\} \right\rangle _{f}-\left\langle \hat{X}_{s}\right\rangle
_{f}\left\langle \hat{X}_{t}\right\rangle _{f}\ .%
\end{array}
\label{35}
\end{equation}%
It is easy to see that $\Delta _{ss}^{PP}=\left\langle \hat{P}%
_{s}^{2}\right\rangle _{f}-\left\langle \hat{P}_{s}\right\rangle
_{f}^{2}=\sigma _{\hat{P}_{s}}^{2}$, $\Delta _{ss}^{XX}=\left\langle \hat{X}%
_{s}^{2}\right\rangle _{f}-\left\langle \hat{X}_{s}\right\rangle
_{f}^{2}=\sigma _{\hat{X}_{s}}^{2}$, and $\Delta _{st}^{XP}=\Delta
_{ts}^{PX} $.

The UR between $\hat{P}_{K_{1},i}$ and $\hat{P}_{K_{2},j}$ for KT that
defined by Eq. (\ref{31}) is now%
\begin{equation}
\sigma _{\hat{P}_{K_{1},i}}^{2}\sigma _{\hat{P}_{K_{2},j}}^{2}\geqslant
F_{K_{1}K_{2},ij}^{2}+\frac{1}{4}W_{K_{1}K_{2},ij}^{2}.  \label{36}
\end{equation}%
For such kind of KT, the UR\ can be derived routinely. First we read off the
parameters $\tilde{B}_{K}$ and $C_{K}$ directly from the $c$-number kernel
Eq. (\ref{31}). Then we do a little bit algebra to calculate $c$-number
matrices $W_{K_{1}K_{2}}\equiv \left( C_{K_{1}}^{T}\right) ^{-1}\left(
\tilde{B}_{K_{1}}-\tilde{B}_{K_{2}}\right) C_{K_{2}}^{-1}$ and $%
F_{K_{1}K_{2}}\equiv \left( C_{K_{1}}^{T}\right) ^{-1}\left( \tilde{B}%
_{K_{1}}\Delta ^{XP}+\Delta ^{PX}\tilde{B}_{K_{2}}+\tilde{B}_{K_{1}}\Delta
^{XX}\tilde{B}_{K_{2}}+\Delta ^{PP}\right) C_{K_{2}}^{-1}$. And this
completes the calculation of UR (\ref{36}).

\section{Some Examples}

In this section we apply the results in last section to four examples: the
FrFT, one dimensional LCT, \ the fractional squeezing transform in \cite%
{CHEN} and the generalized time-frequency transform in \cite{Sahay}. The
first two examples have been calculated in \cite{ITE1,ITE2,ITE3,ITE4}. The
new method we apply here gives the same results, but with much shorter
length and less efforts. The difficult parts of calculation have been
completed in the previous sections and generic results (\ref{25}) and (\ref%
{36}) have been derived. What we need to do in the following is no more than
plug-in-the-parameters for each case. The third and the fourth ones are new
transformations. The calculations are also simple and straightforward.

\subsection{Traditional FrFT}

As is well-known, the 1-D FrFT kernel is
\begin{equation}
K_{\alpha }\left( p,x\right) =\sqrt{\frac{1-i\cot \alpha }{2\pi }}e^{\frac{i%
}{2}\left( \frac{p^{2}+x^{2}}{\tan \alpha }-\frac{2px}{\sin \alpha }\right)
}.  \label{37}
\end{equation}%
Comparing (\ref{37}) with (\ref{31}), we have $C_{\alpha }^{-1}=\sin \alpha $
and$\ \tilde{B}_{\alpha }=\cot \alpha $ here. Then according to Eq. (\ref{32}%
), the transformed momentum is%
\begin{equation}
\hat{p}_{\alpha }=\hat{p}\sin \alpha +\hat{x}\cos \alpha .  \label{38}
\end{equation}%
And
\begin{eqnarray}
\Delta ^{PP} &=&\sigma _{\hat{p}}^{2},\ \Delta ^{XX}=\sigma _{\hat{x}}^{2},
\label{39} \\
\Delta ^{PX} &=&\Delta ^{XP}=\left\langle \frac{1}{2}\left\{ \hat{p},\hat{x}%
\right\} \right\rangle _{f}-\left\langle \hat{p}\right\rangle
_{f}\left\langle \hat{x}\right\rangle _{f}  \notag \\
&\equiv &R_{xp}\sqrt{\sigma _{\hat{x}}^{2}\sigma _{\hat{p}}^{2}},  \notag
\end{eqnarray}%
where
\begin{equation}
R_{xp}=\frac{\frac{1}{2}\left\langle \hat{x}\hat{p}+\hat{p}\hat{x}%
\right\rangle -\bar{x}\bar{p}}{\sqrt{\sigma _{\hat{x}}^{2}\sigma _{\hat{p}%
}^{2}}}  \label{40}
\end{equation}%
is the correlation coefficient between observables $\hat{x}$\ and $\hat{p}$.
Eqs. (\ref{33}, \ref{34}) become
\begin{eqnarray}
W_{\alpha \beta } &=&\sin \left( \beta -\alpha \right) ,  \label{41} \\
F_{\alpha \beta } &=&\sigma _{\hat{p}}^{2}\sin \alpha \sin \beta +\sigma _{%
\hat{x}}^{2}\cos \alpha \cos \beta  \notag \\
&&+R_{xp}\sqrt{\sigma _{\hat{x}}^{2}\sigma _{\hat{p}}^{2}}\sin \left( \alpha
+\beta \right) .  \notag
\end{eqnarray}%
The UR Eq. (\ref{36}) for 1-D FrFT reads
\begin{eqnarray}
\sigma _{\hat{p}_{\alpha }}^{2}\sigma _{\hat{p}_{\beta }}^{2} &\geqslant
&\left\vert
\begin{array}{c}
\sigma _{\hat{p}}^{2}\sin \alpha \sin \beta \\
+R_{xp}\sqrt{\sigma _{\hat{x}}^{2}\sigma _{\hat{p}}^{2}}\sin \left( \alpha
+\beta \right) \\
+\sigma _{\hat{x}}^{2}\cos \alpha \cos \beta%
\end{array}%
\right\vert ^{2}  \label{42} \\
&&+\frac{1}{4}\sin ^{2}\left( \alpha -\beta \right) .  \notag
\end{eqnarray}%
In the case that $f\left( x\right) =\left\langle x\right. \left\vert
f\right\rangle $ is a real function (real signal), we see
\begin{eqnarray}
\left\langle \hat{x}\hat{p}\right\rangle _{f} &=&\int_{-\infty }^{\infty
}f\left( x\right) x\frac{1}{i}\frac{d}{dx}f\left( x\right) dx  \label{43} \\
&=&-\frac{i}{2}\int_{-\infty }^{\infty }xd\left[ f^{2}\left( x\right) \right]
\notag \\
&=&xf^{2}\left( x\right) |_{-\infty }^{\infty }+\frac{i}{2}\int_{-\infty
}^{\infty }f^{2}\left( x\right) dx  \notag \\
&=&\frac{i}{2},  \notag
\end{eqnarray}%
and $\left\langle \hat{p}\hat{x}\right\rangle _{f}=\left\langle \hat{x}\hat{p%
}\right\rangle _{f}^{\ast }=-\frac{i}{2}$. We have
\begin{eqnarray}
\sigma _{p_{\alpha }}^{2}\sigma _{p_{\beta }}^{2} &\geqslant &\left\vert
\begin{array}{c}
\sigma _{x}^{2}\cos \alpha \cos \beta +\sigma _{p}^{2}\sin \alpha \sin \beta
\\
-\bar{x}\bar{p}\sin \left( \alpha +\beta \right)%
\end{array}%
\right\vert ^{2}  \label{44} \\
&&+\frac{1}{4}\sin ^{2}\left( \alpha -\beta \right) .  \notag
\end{eqnarray}%
Further, when $\bar{x}=0$, then%
\begin{eqnarray}
\sigma _{p_{\alpha }}^{2}\sigma _{p_{\beta }}^{2} &\geqslant &\left\vert
\sigma _{x}^{2}\cos \alpha \cos \beta +\sigma _{p}^{2}\sin \alpha \sin \beta
\right\vert ^{2}  \label{45} \\
&&+\frac{1}{4}\sin ^{2}\left( \alpha -\beta \right) .  \notag
\end{eqnarray}%
At this point we mention that Eq. (24) in \cite{ITE1} can re-appear here
directly from Eq. (\ref{45}).

Particularly, when $R_{xp}=0$ and $\beta =0$, Eq. (\ref{42}) becomes%
\begin{equation}
\sigma _{x}^{2}\sigma _{p_{\alpha }}^{2}\geqslant \left( \sigma
_{x}^{2}\right) ^{2}\cos ^{2}\alpha +\frac{1}{4}\sin ^{2}\alpha ,  \label{46}
\end{equation}%
this is exactly Eq. (49) of \cite{ITE1}.

\subsection{One Dimensional LCT}

The kernel for one dimensional LCT with parameter $M=\left( a,b,c,d\right) $
is

\begin{equation}
K_{M}\left( p,x\right) =\sqrt{\frac{1}{2\pi ib}}e^{i\left( \frac{d}{2b}p^{2}+%
\frac{a}{2b}x^{2}-\frac{px}{b}\right) }.  \label{47}
\end{equation}%
Comparing (\ref{47}) with (\ref{31}), we have $C_{M}^{-1}=b$ and$\ \tilde{B}%
_{M}=\frac{a}{b}$ here. Then according to Eq. (\ref{32}), the transformed
momentum is

\begin{equation}
\hat{p}_{M}=b\hat{p}+a\hat{x}.  \label{48}
\end{equation}%
And%
\begin{eqnarray}
W_{M_{1}M_{2}} &=&a_{1}b_{2}-a_{2}b_{1},  \label{49} \\
F_{M_{1}M_{2}} &=&a_{1}a_{2}\sigma _{\hat{x}}^{2}+b_{1}b_{2}\sigma _{\hat{p}%
}^{2}  \notag \\
&&+\left( a_{1}b_{2}+a_{2}b_{1}\right) Cov\left( x,p\right) .  \notag
\end{eqnarray}%
The uncertainty relation%
\begin{eqnarray}
\sigma _{\hat{P}_{M_{1}}}^{2}\sigma _{\hat{P}_{M_{2}}}^{2} &\geqslant
&F_{M_{1}M_{2}}^{2}+\frac{1}{4}W_{M_{1}M_{2}}^{2}  \label{50} \\
&=&\left(
\begin{array}{c}
a_{1}a_{2}\sigma _{\hat{x}}^{2}+b_{1}b_{2}\sigma _{\hat{p}}^{2} \\
+\left( a_{1}b_{2}+a_{2}b_{1}\right) Cov\left( x,p\right)%
\end{array}%
\right) ^{2}  \notag \\
&&+\frac{1}{4}\left( a_{1}b_{2}-a_{2}b_{1}\right) ^{2}.  \notag
\end{eqnarray}%
obtained here is exactly the main result Eq. (20) in \cite{ITE4}. And the
method we used here can be applied easily to LCT of any dimensions.

\subsection{Fractional Squeezing Transform}

As the second example in \cite{CHEN}, still we take%
\begin{equation}
\begin{array}{c}
\left\vert \boldsymbol{A}\right\rangle =\left\vert x\right\rangle =\frac{1}{%
\pi ^{1/4}}\exp \left[ -\frac{1}{2}x^{2}+\sqrt{2}xa^{\dag }-\frac{1}{2}%
a^{\dag 2}\right] \left\vert 0\right\rangle , \\
\left\vert \boldsymbol{B}\right\rangle =\left\vert p\right\rangle =\frac{1}{%
\pi ^{1/4}}\exp \left[ -\frac{1}{2}p^{2}+i\sqrt{2}pa^{\dag }+\frac{1}{2}%
a^{\dag 2}\right] \left\vert 0\right\rangle ,%
\end{array}
\label{51}
\end{equation}%
And we introduce the fractional squeezing transform defined by the operator%
\begin{equation}
\mathbf{\hat{K}}_{\alpha }=\exp \left[ -\frac{i\alpha }{2}\left(
a^{2}e^{i\theta }+e^{-i\theta }a^{\dag 2}\right) \right] \exp \left[ \frac{%
i\pi }{2}a^{\dag }a\right] .  \label{52}
\end{equation}%
The $c$-number kernel for the fractional squeezing transform is%
\begin{equation}
K_{\alpha }\left( p,x\right) =\frac{\exp \left[ \frac{i}{2}\left( \frac{%
x^{2}+p^{2}}{\tanh \alpha \cos \theta }+\frac{x^{2}-p^{2}}{\cot \theta }-%
\frac{2xp}{\sinh \alpha \cos \theta }\right) \right] }{\sqrt{2\pi i\cos
\theta \sinh \alpha }}.  \label{53}
\end{equation}%
The fractional squeezing transform is additive, $T_{\alpha }\circ T_{\beta
}=T_{\alpha +\beta }$.

Comparing (\ref{53}) with (\ref{31}), we have%
\begin{equation}
C_{\alpha }^{-1}=\sinh \alpha \cos \theta ,\ \tilde{B}_{\alpha }=\frac{%
1+\tanh \alpha \sin \theta }{\tanh \alpha \cos \theta }.  \label{54}
\end{equation}%
According to Eq. (\ref{32}), the transformed momentum is%
\begin{equation}
\hat{p}_{\alpha }=\hat{p}\sinh \alpha \cos \theta +\hat{x}\left( \cosh
\alpha +\sinh \alpha \sin \theta \right) .  \label{55}
\end{equation}%
Eqs. (\ref{33}, \ref{34}) become%
\begin{equation}
\begin{array}{c}
W_{\alpha \beta }=\cos \theta \sinh \left( \beta -\alpha \right) , \\
F_{\alpha \beta }=\sigma _{\hat{p}}^{2}\sinh \alpha \sinh \beta \cos
^{2}\theta \\
+R_{xp}\sqrt{\sigma _{\hat{x}}^{2}\sigma _{\hat{p}}^{2}}\left[ \sinh \left(
\alpha +\beta \right) +\sinh \alpha \sinh \beta \sin 2\theta \right] \\
+\sigma _{\hat{x}}^{2}\left( \cosh \alpha +\sinh \alpha \sin \theta \right)
\left( \cosh \beta +\sinh \beta \sin \theta \right) .%
\end{array}
\label{56}
\end{equation}%
The UR of signals on the fractional squeezing transform is%
\begin{equation}
\begin{array}{c}
\sigma _{p_{\alpha }}^{2}\sigma _{p_{\beta }}^{2}\geqslant \left\vert
\begin{array}{c}
\sigma _{\hat{p}}^{2}\sinh \alpha \sinh \beta \cos ^{2}\theta \\
+\sigma _{\hat{x}}^{2}\left( \cosh \alpha +\sinh \alpha \sin \theta \right)
\times \\
\times \left( \cosh \beta +\sinh \beta \sin \theta \right) \\
+R_{xp}\sqrt{\sigma _{\hat{x}}^{2}\sigma _{\hat{p}}^{2}}\left[
\begin{array}{c}
\sinh \alpha \sinh \beta \sin 2\theta \\
+\sinh \left( \alpha +\beta \right)%
\end{array}%
\right]%
\end{array}%
\right\vert ^{2} \\
+\frac{1}{4}\cos ^{2}\theta \sinh ^{2}\left( \alpha -\beta \right) .%
\end{array}
\label{57}
\end{equation}

The fractional squeezing transform is a new GFrT, which is additive just
like FrFT. We believe it will be widely used in physics, data analysis and
signal processing.

\subsection{The Generalized Time-Frequency Transform in \protect\cite{Sahay}}

In \cite{Sahay}, Sahay et al proposed a new KT defined by kernel (Eq. 45 in
\cite{Sahay})%
\begin{equation}
K_{\phi ,\psi }\left( x,p\right) =\sqrt{\frac{1}{2\pi il\left( \phi \right) }%
}\exp \left[ i\left( p^{2}+x^{2}\right) g\left( \phi \right) -\frac{ixp}{%
l\left( \phi \right) }+if\left( p,\psi \right) -if\left( x,\psi \right) %
\right]  \label{58}
\end{equation}

Although it looks complicated and contains arbitrary parameters $g\left(
\phi \right) $, $l\left( \phi \right) $ and function $f\left( p,\psi \right)
$, this kernel still takes the form of Eq. (\ref{28}). Therefore according
to Eq. (\ref{32}), the transformed momentum operator is%
\begin{eqnarray}
\hat{p}_{\phi ,\psi } &=&\left( C_{K}^{T}\right) ^{-1}\left( \hat{p}+\nabla
_{\hat{X}}B_{K}\left( \hat{X}\right) \right)  \label{59} \\
&=&l\left( \phi \right) \left[ \hat{p}+2\hat{X}g\left( \phi \right) -\frac{%
\partial }{\partial \hat{X}}f\left( \hat{X},\psi \right) \right]  \notag
\end{eqnarray}%
Plug Eq. (\ref{59}) into Eq. (\ref{24}) we will get the UR for this
generalized time-frequency transform.

As an example, let us choose $l\left( \phi \right) =\sin \phi $, $g\left(
\phi \right) =\frac{1}{2}\cot \phi $ and $f\left( x,\psi \right) =x^{3}$ as
in Eq. 44 of \cite{Sahay}, then%
\begin{equation}
\hat{p}_{\phi }=\hat{p}\sin \phi +\hat{X}\cos \phi -3\hat{X}^{2}\sin \phi .
\label{60}
\end{equation}%
Therefore%
\begin{equation}
\frac{1}{i}\left[ \hat{p}_{\phi _{1}},\hat{p}_{\phi _{2}}\right] =\sin
\left( \phi _{2}-\phi _{1}\right) .  \label{61}
\end{equation}%
And the covariance of $\hat{p}_{\phi _{1}}$, $\hat{p}_{\phi _{2}}$ is
\begin{eqnarray}
&&Cov\left( \hat{p}_{\phi _{1}},\hat{p}_{\phi _{2}}\right)   \label{62} \\
&=&\frac{1}{2}\left\langle \left\{ \hat{p}_{\phi _{1}},\hat{p}_{\phi
_{2}}\right\} \right\rangle _{h}-\left\langle \hat{p}_{\phi
_{1}}\right\rangle _{h}\left\langle \hat{p}_{\phi _{2}}\right\rangle _{h}
\notag \\
&=&\left( \sigma _{\hat{p}}^{2}-3Cov\left( \hat{x}^{2},\hat{p}\right)
+9\sigma _{\hat{x}^{2}}^{2}\right) \sin \phi _{1}\sin \phi _{2}+\sigma _{%
\hat{x}}^{2}\cos \phi _{1}\cos \phi _{2}  \notag \\
&&+\left[ Cov\left( \hat{x},\hat{p}\right) -3Cov\left( \hat{x},\hat{x}%
^{2}\right) \right] \sin \left( \phi _{1}+\phi _{2}\right)   \notag
\end{eqnarray}%
where%
\begin{eqnarray}
\sigma _{\hat{p}}^{2} &=&\left\langle \hat{p}^{2}\right\rangle
_{h}-\left\langle \hat{p}\right\rangle _{h}^{2}  \label{63} \\
\sigma _{\hat{x}}^{2} &=&\left\langle \hat{x}^{2}\right\rangle
_{h}-\left\langle \hat{x}\right\rangle _{h}^{2}  \notag \\
\sigma _{\hat{x}^{2}}^{2} &=&\left\langle \hat{x}^{4}\right\rangle
_{h}-\left\langle \hat{x}^{2}\right\rangle _{h}^{2}  \notag \\
Cov\left( \hat{x}^{2},\hat{p}\right)  &=&\frac{1}{2}\left\langle \left\{
\hat{x}^{2},\hat{p}\right\} \right\rangle _{h}-\left\langle \hat{x}%
^{2}\right\rangle _{h}\left\langle \hat{p}\right\rangle _{h}  \notag \\
Cov\left( \hat{x},\hat{p}\right)  &=&\frac{1}{2}\left\langle \left\{ \hat{x},%
\hat{p}\right\} \right\rangle _{h}-\left\langle \hat{x}\right\rangle
_{h}\left\langle \hat{p}\right\rangle _{h}  \notag \\
Cov\left( \hat{x},\hat{x}^{2}\right)  &=&\left\langle \hat{x}%
^{3}\right\rangle _{h}-\left\langle \hat{x}\right\rangle _{h}\left\langle
\hat{x}^{2}\right\rangle _{h}  \notag
\end{eqnarray}%
Finally we have the new UR for this generalized time-frequency transform.
\begin{equation}
\sigma _{\hat{p}_{\phi _{1}}}^{2}\sigma _{\hat{p}_{\phi _{2}}}^{2}\geqslant
\sin ^{2}\left( \phi _{1}-\phi _{2}\right) +\left\vert Cov\left( \hat{p}%
_{\phi _{1}},\hat{p}_{\phi _{2}}\right) \right\vert ^{2}  \label{64}
\end{equation}

\section{Conclusion}

In summary, we have derived the generic uncertainty relation Eq. (\ref{25})
for kernel-based transformations. And explicit UR's Eq. (\ref{36}) are
derived for a family of KT's including GFrT and multi-dimensional LCT.
Instead of using the statistics approach for FrFT and LCT as shown in the
literatures before, which takes tedious work, here we have employed quantum
mechanical operator approach for directly deriving the UR for KT. We are
able to do this because we have found the quantum operator realization of
KT. Our new method is concise and applicable to any kinds of KTs, with
continuous and discrete parameters and variables.

\section*{Acknowledgment}

Work supported by the National Natural Science Foundation of China under
grant: 11105133 and 11175113, and National Basic Research Program of China
(973 Program, 2012CB922001).


\begin{thebibliography}{99}
\bibitem{1} E. U. Condon, \emph{Immersion of the Fourier transform in a
continuous group of functional transformations}, Proc. Natl. Acad. Sci. USA,
Vol. 23, No. 3, pp158-164, 1937.

\bibitem{2} V. Namias, \emph{The fractional Fourier transform and its
application in quantum mechanics}, J. Inst. Math. Its Appl., Vol. 25, No. 3,
pp241-265, 1980.

\bibitem{3} D.~Mendlovic and H. M. Ozaktas, \emph{Fractional Fourier
transforms and their optical implementation, I}, J. Opt. Soc. Am. A, Vol.
10, No. 9, pp1875-1881, 1993.

\bibitem{4} D. Mendlovic, H. M. Ozaktas and A. W. Lohmmann, \emph{%
Graded-index fiber, Wigner-distribution functions, and the fractional
Fourier transform}, Appl. Opt., Vol. 33, No. 26, pp6188-6193, 1994.

\bibitem{5} A. C. McBride and F. H. Kerr, \emph{On Namias' fractional
Fourier transform}, IMA J. Appl. Math., Vol. 39, No. 2, pp159-175, 1987.

\bibitem{6} A. W. Lohmann, \emph{Image rotation, Wigner rotation and
fractional Fourier transform}, J. Opt. Soc. Am. A, Vol. 10, No. 10,
2181-2186, 1993.

\bibitem{7} L. Bernardo and O. D. D. Soares, \emph{Fractional Fourier
transform and optical systems}, Opt. Commun., Vol. 110, No. 5-6, pp517-522,
1994.

\bibitem{JPA} Sumiyoshi Abet and John T Sheridant, \emph{Generalization of
the fractional Fourier transformation to an arbitrary linear lossless
transformation: an operator approach}, J. Phys. A: Math. Gen., Vol. 27, No.
12, pp4179-4187, 1994.

\bibitem{CHEN} Jun-Hua Chen and Hong-Yi Fan, \emph{Quantum mechanical
perspectives and generalization of the fractional Fourier transformation},
arXiv:1307.6271 [math-ph]

\bibitem{JMP1} Meng-Sen Ma and Ren Zhao, \emph{The effect of generalized
uncertainty principle on square well, a case study}, Journal of Mathematical
Physics, Vol. 55, No. 8, pp082109, 2014

\bibitem{JMP2} J. Crann and M. Kalantar, \emph{An uncertainty principle for
unimodular quantum groups}, Journal of Mathematical Physics, Vol. 55, No. 8,
pp081704, 2014

\bibitem{JMP3} Yan-Ni Dou and Hong-Ke Du, \emph{Generalizations of the
Heisenberg and Schrodinger uncertainty relations}, Journal of Mathematical
Physics, Vol. 54, No. 10, pp103508, 2014

\bibitem{JMP4} Vasily E. Tarasov, \emph{Uncertainty relation for
non-Hamiltonian quantum systems}, Journal of Mathematical Physics, Vol. 54,
No. 1, pp012112, 2013

\bibitem{ITE1} Sudarshan Shinde and Vikram M. Gadre, \emph{An Uncertainty
Principle for Real Signals in the Fractional Fourier Transform Domain}, IEEE
Transactions on Signal Processing, Vol. 49, No. 11, pp2545-2548, 2001.

\bibitem{ITE2} K. K. Sharma and S. D. Joshi, \emph{Uncertainty Principle for
Real Signals in the Linear Canonical Transform Domains}, IEEE Transactions
on Signal Processing, Vol. 56, No. 7, pp2677-2683, 2008.

\bibitem{ITE3} Guanlei Xu, Xiaotong Wang and Xiaogang Xu, \emph{On
uncertainty principle for the linear canonical transform of complex signals}%
, IEEE Transactions on Signal Processing, Vol. 58, No. 9, pp4916-4918, 2010.

\bibitem{ITE4} J. Zhao, R. Tao, and Y. Wang, \emph{On signal moments and
uncertainty relations associated with linear canonical transform}, IEEE
Transactions on Signal Processing, Vol. 90, No. 9, pp2686-2689, 2010.

\bibitem{Robertson} H. P. Robertson, \emph{The Uncertainty Principle}, Phys.
Rev. Vol. 34 pp163-164, 1929.

\bibitem{Schrodinger} E. Schr\"{o}dinger, \emph{Zum Heisenbergschen Unsch%
\"{a}rfeprinzip}, Sitzungsberichte der Preussischen Akademie der
Wissenschaften, Physikalisch-mathematische Klasse Vol. 14, pp296-303, 1930.

\bibitem{Sahay} S. Sahay, D. Pande, V. Gadre and P. Sohani, \emph{A Novel
Generalized Time-Frequency Transform Inspired by the Fractional Fourier
Transform for Higher Order Chirps}, Signal Processing and Communications
(SPCOM), 2012 International Conference on, DOI: 10.1109/SPCOM.2012.6289994
\end{thebibliography}
\end{document}